\documentclass[runningheads,a4paper,oribibl]{llncs}
\usepackage[english]{babel}
\usepackage{bussproofs}
\usepackage{graphicx,alltt}
\usepackage{hyperref}
\hypersetup{
    colorlinks=true,
    citecolor=blue,
    linkcolor=blue,
    urlcolor=blue,
}
\usepackage{color}
\usepackage{listings}

\definecolor{gray}{rgb}{0.3,0.3,0.3} 
\definecolor{dkblue}{rgb}{0,0.1,0.5} 
\definecolor{lightblue}{rgb}{0,0.5,0.5} 
\definecolor{dkgreen}{rgb}{0,0.4,0} 
\definecolor{dk2green}{rgb}{0.4,0,0} 
\definecolor{dkviolet}{rgb}{0.6,0,0.8}
\definecolor{dkpink}{rgb}{0.8,0,0.9}
\newcommand{\tmverbatim}[1]{\lstinline|#1|}

\begin{document}

\title{Asynchronous~processing~of~Coq~documents: from the kernel up to the user interface}
\titlerunning{Asynchronous~processing~of~Coq~documents}

\author{ 
	 Bruno Barras
	 \and
         Carst Tankink 
	 \and 
	 Enrico Tassi
}

\institute{Inria\\
~\\
\email{
 \{bruno.barras,
    carst.tankink,
    enrico.tassi%
 \}@inria.fr}
}

\maketitle

\begin{abstract}
  The work described in this paper improves the reactivity of the Coq
  system by completely redesigning the way it processes a formal
  document.  By subdividing such work into independent
  tasks the system can give precedence to the ones of immediate
  interest for the user and postpone the others.  On the user side, a
  modern interface based on the PIDE middleware aggregates and
  presents in a consistent way the output of the prover.  Finally
  postponed tasks are processed exploiting modern, parallel,
  hardware to offer better scalability.
\end{abstract}

\section{Introduction}

In recent years Interactive Theorem Provers (ITPs) have been
successfully used to give an ultimate degree of reliability to both
complex software systems, like the L4 micro
kernel~\cite{Klein:2010:SFV:1743546.1743574} and the CompCert C
compiler~\cite{Leroy-Compcert-CACM}, and mathematical theorems, like
the Odd Order Theorem~{\cite{gonthier:hal-00816699}}.  These large
formalization projects have pushed interactive provers to their
limits, making their deficiencies apparent.
The one we deal with in this work is {\emph{reactivity}}: how long it
takes for the system to react to a user change and give her feedback on
her point of interest. For example if one takes the full proof of the
Odd Order Theorem, makes a change in the first file and asks the
Coq prover for any kind of feedback on the last file she has to wait
approximately two hours before receiving any feedback. 

To find a solution to this problem it is important to understand how
formal documents are edited by the user and checked by the prover.
Historically ITPs have come with a simple text based Read Eval Print
Loop (REPL) interface: the user inputs a command, the system runs it
and prints a report.  It is up to the user to stock the right sequence
of commands, called script, in a file. The natural evolution
of REPL user interfaces adds a very basic form of script
management on top of a text editor: the user writes his commands
inside a text buffer and tells the User Interface (UI) to send them one by
one to the same REPL interface.  This
design 
is so cheap, in terms of
coding work required on the prover side, and so generic that its best
incarnation, Proof General~{\cite{DBLP:conf/tacas/Aspinall00}}, has
served as the reference UI for many provers, Coq included, for over a
decade.

The simplicity of REPL comes at a price: commands must be executed in a linear
way, one after the other. For example the prover must tell the UI if
a command fails or succeeds before the following command can be sent
to the system.  Under such constraint to achieve better reactivity one
needs to speed up the execution of each and every command composing
the formal document.  Today one would probably do that by taking
advantage of modern, parallel hardware.  Unfortunately it is very hard
to take an existing system coded in a imperative style and parallelize
its code at the fine grained level of single commands.  Even more if
the programming language it is written in does not provide light
weight execution threads.  Both conditions apply to Coq.

If we drop the constraint imposed by the REPL, a different,
complementary, way to achieve better reactivity becomes an option:
process the formal document \emph{out-of-order},\footnote{In analogy
with the ``out-of-order execution'' paradigm used in
most high-performance microprocessors} giving precedence to
the parts the user is really interested in and postpone the others.
In this view, even if commands do not execute faster, the system needs
to execute fewer of them in order to give feedback to the user.  In
addition to that, when the parts the document is split in happen to be
independent, the system can even process them in parallel.

Three ingredients are crucial to process a formal document out-of-order.

First, the UI must not impose to the prover to run commands linearly.
A solution to this problem has been studied by Wenzel
in~\cite{DBLP:journals/corr/Wenzel13} for the Isabelle
system. His approach consists in
replacing the REPL with an \emph{asynchronous interaction loop}: each
command is marked with a unique identifier and each report generated
by the prover carries the identifier of the command to which it applies.
The user
interface sends the whole document to the prover and uses these unique
identifiers to 
present the prover outputs coherently to the
user.  The asynchronous interaction loop developed by Wenzel is part
of a generic middleware called PIDE (for Prover IDE) that we extend
with a Coq specific back end.

Second, the prover must be able to perform a \emph{static analysis of
the document} in order to organize it into coarse grained tasks and
take scheduling decisions driven by the user's point of
interest.  Typically a task is composed of many consecutive commands.
In the case of Coq a task corresponds to the sequence of tactics,
proof commands, that builds an entire proof.  In other words the text between
the \tmverbatim{Proof} and \tmverbatim{Qed} keywords.

Third, the system must feature an \emph{execution model} that allows
the reordering of tasks.  In particular we model tasks as pure
computations and we analyze the role their output plays in the checking
of the document. Finally we implement the machinery required in order
to execute them in parallel by using the coarse grained concurrency
provided by operating system processes.

In this work we completely redesigned from the ground up
the way Coq processes a formal document
in order to obtain the three
ingredients above.  The result is a more reactive system that also
performs better at checking documents in batch mode.  Benchmarks show
that the obtained system is ten times more reactive when processing
the full proof of the Odd Order Theorem and that it scales better when
used on parallel hardware.  In particular fully checking such proof on
a twelve core machine is now four times faster than before.  Finally,
by plugging Coq in the PIDE middleware we get a modern user interface
based on the jEdit editor that follows the ``spell checker paradigm''
made popular by tools like Eclipse or Visual Studio: the user
freely edits the document and the prover constantly annotates it with
its reports, like underlining in red problematic sentences.
The work of
Wenzel~\cite{Wenzel_parallelproof,DBLP:conf/mkm/Wenzel11,DBLP:journals/corr/Wenzel13,Wenzel14} on the
Isabelle system has laid the foundations for our design, and has been
a great inspiration for this work.  The design and implementation work
spanned over three years and the results are part of version 8.5 of
the Coq system.  All authors were supported by the Paral-ITP
ANR-11-INSE-001 project.

The paper is organized as follows.  Section~\ref{document} describes
the main data structure used by the prover in order to statically
analyze the document and represent the resulting tasks.
Section~\ref{futures} describes how asynchronous, parallel,
computations are modeled in the logical kernel of the system and how
they are implemented in the OCaml programming language.
Section~\ref{pide} describes how the document is represented on the UI
side, how data is aggregated and presented to the user.
Section~\ref{12core} presents a reactivity benchmark of the redesigned
system on the full proof of the Odd Order Theorem.
Section~\ref{conclusions} 
concludes.

%

\section{Processing the formal document out-of-order}\label{document}

Unlike REPL in the asynchronously interaction model promoted by
PIDE~\cite{DBLP:journals/corr/Wenzel13} 
the prover is made aware of the whole document and it is expected to process
it giving precedence to the portion of the text the user is looking at. To do
so the system must identify the parts of the document that are not relevant to
the user and postpone their processing. The most favorable case is when large
portions of the document are completely independent from each other, and hence
one has complete freedom on the order in which they must be processed. In the specific
case of Coq, {\emph{opaque proofs}} have this property. Opaque proofs are the
part of the document where the user builds a proof evidence (a proof
term in
Coq's terminology) by writing a sequence of tactics and that ends with the
\tmverbatim{Qed} keyword (lines four to seven in Figure~\ref{dag}). The
generated proof term is said to be opaque because it is verified
by the kernel of the system and stored on disk, but the term is never
used, only its corresponding statement (its type) is. The user can ask the
system to print such term or to translate it to OCaml code, but from the
viewpoint of the logic of Coq (type theory) the system commits not to use the
proof term while checking other proofs.{\footnote{In the Curry-Howard
correspondence Coq builds upon lemmas and definitions are the same: a
term of a given type. An opaque lemma is a definition one cannot unfold.}}

The notion of proof opacity was introduced a long ago in Coq version 5.10.5
(released in May 1994) and is crucial for us: given that the proof term is not
used,
we can postpone the processing of the piece of the
document that builds it as much as we want. All we need is its type, that is
the statement that is spelled out explicitly by the user. Proofs are also
lengthy and usually the time spent in processing them dominates the overall
time needed in order to check the entire document. For example in the 
case of the Odd Order Theorem proofs amount to 60\% of the non-blanks
(3.175KB over 5.262KB) and Coq spends 90\% of its time on them. This means
that, by properly scheduling the processing of opaque proofs, we can increase
the reactivity of the system of a factor of ten. In addition to that, the
independence of each proof from the others makes it possible to process many
proofs at the same time, in parallel, giving a pretty good occasion to exploit
modern parallel hardware.

In the light of that, it is crucial to build an internal representation for
the document that makes it easy to identify opaque proofs.
Prior to this work Coq had no
internal representation of the document at all. To implement the
\tmverbatim{Undo} facility, Coq has a notion of {\emph{system state}} that
can can be saved and restored on demand. But the system used to keep no trace
of how a particular system state was obtained; which sequence of commands
results in a particular system state.

The most natural data structure for keeping track of how a system state is
obtained is a Directed Acyclic Graph (DAG), where nodes are system states and
edges are labeled with commands. The general idea is that in order to obtain a
system state from another one, one has to process all the commands on the edges
linking the two states. The software component in charge of building and
maintaining such data structure is called State Transaction Machine (STM),
where the word transaction is chosen to stress that commands need to be
executed atomically: there is no representation for a partially
executed command.

\subsection{The STM and the static analysis of the document}\label{static}

\begin{figure}[!tb]
\begin{lstlisting}
1 (* global *)    Definition decidable (P : Prop) := P \/ ~ P.
2 
3 (* branch *)    Theorem dec_False : decidable False.
4 (* tactic *)    Proof.
5 (* tactic *)     unfold decidable, not.
6 (* tactic *)     auto.
7 (* merge $~$*)    Qed.
\end{lstlisting}
\begin{center}
\includegraphics[width=0.9\textwidth]{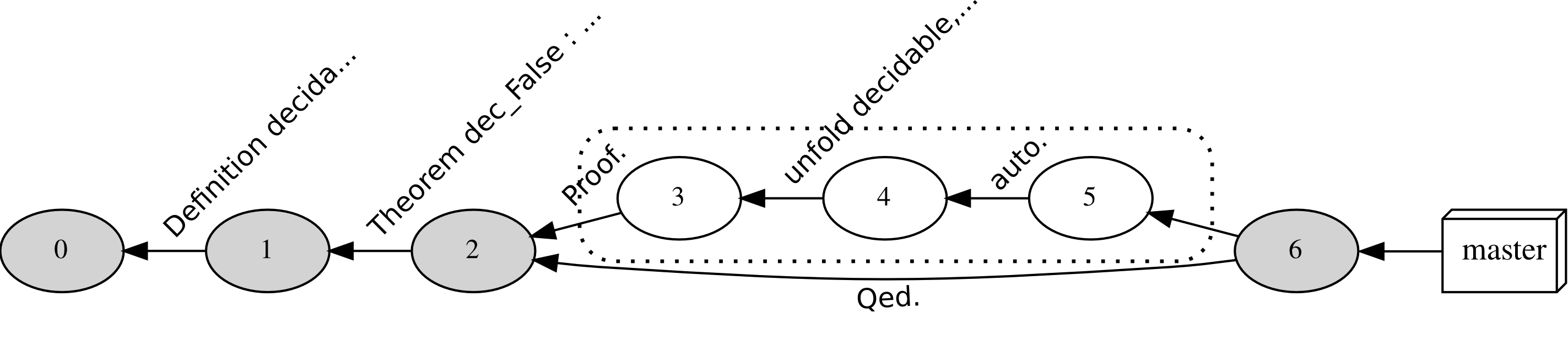}
\end{center}
\caption{\label{dag}Coq document and its internal representation}
\end{figure}

The goal of the static analysis of the document is to build a DAG in which
proofs are explicitly separated from the rest of the document. We first parse
each sentence
obtaining the abstract syntax tree of the corresponding
command in order to classify it. Each command belongs to only one of the
following categories: commands that start a proof ({\emph{branch}}), commands
that end a proof ({\emph{merge}}), proof commands ({\emph{tactic}}), and
commands that have a global effect ({\emph{global}}). The DAG is built in the
very same way one builds the history of a software project in a version
control system. One starts with an initial state and a default branch, called
master, and proceeds by adding new commits for each command. A commit is a new
node and an edge pointing to the previous node. Global commands add the commit
on the main branch; branching commands start a new branch; tactic commands add
a commit to the current branch; merging commands close the current branch by
merging it into master. If we apply these simple rules to the document in
Figure~\ref{dag} we obtain a DAG where the commands composing the proof of
\tmverbatim{dec\_False} have been isolated. Each node is equipped with a unique
identifier, here positive numbers. The edge from the state six to state five
has no label, it plays the role of making the end of the proof easily
accessible but has not to be ``followed'' when generating state six.

Indeed to compute state six, or anything that follows it,
Coq starts from the initial state, zero, and then executes the labeled
transactions until it reaches state six, namely
``\tmverbatim{Definition}\ldots'',
``\tmverbatim{Theorem}\ldots'' and \tmverbatim{Qed}.
The nodes in gray (1, 2 and 6) are the ones whose
corresponding system state has been computed. The nodes and transactions in
the dotted region compose the proof that is processed asynchronously. 
As a consequence of that the implementation of the
merging command has to be able to accommodate the situation where the proof
term has not been produced yet. This is the subject of Section~\ref{futures}.
For now the only relevant characteristic of the asynchronous processing of
opaque proofs is that such process is a pure computation.
The result of a pure computation does depend only on its input. It
does not
matter when it
is run nor in which environment and it has no visible global effect. If we are
not immediately interested in its result we can execute it lazily,
in parallel or even remotely via the network.

\begin{figure}[!h]
\begin{lstlisting}
1 (* global *)    Definition decidable (P : Prop) := P \/ ~ P.
2 
3 (* branch *)    Theorem dec_False : decidable False.
4 (* tactic *)    Proof.
5 (* global *)     Hint Extern 1 => unfold decidable, not.
6 (* tactic *)     auto.
7 (* merge $~$*)    Qed.
\end{lstlisting}
\begin{center}
\includegraphics[width=0.9\textwidth]{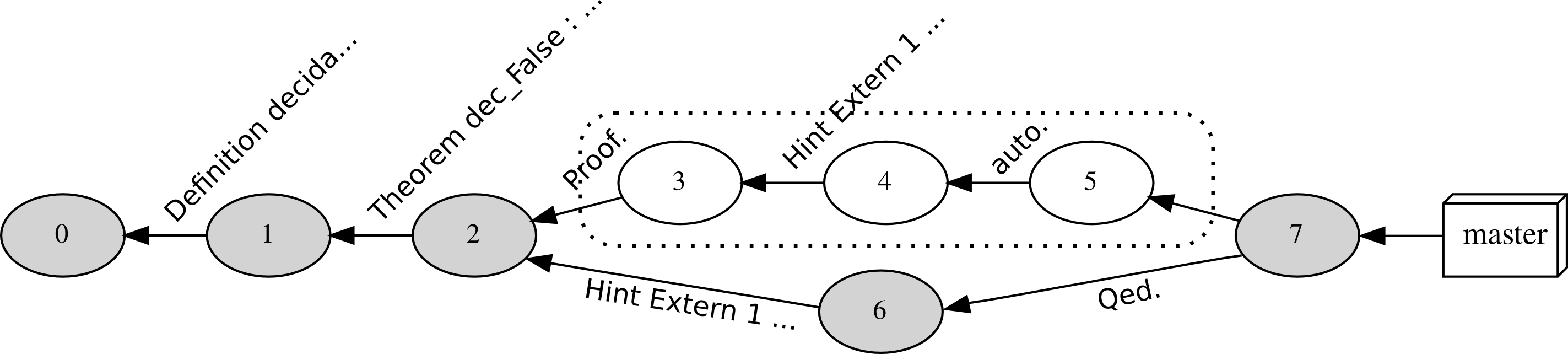}
\end{center}
\caption{\label{dag2}Coq document and its internal representation}
\end{figure}

Tactic commands are not allowed to appear outside a proof; on the contrary
global commands can, an in practice do, appear in the middle of proofs, as in
Figure~\ref{dag2}. Mixing tactics with global commands is not a recommend
style for finished proof scripts, but it is a extremely
common practice while one
writes the script and it must be supported by the system. The current
semantics of Coq documents makes the effect of global commands appearing in
the middle of proofs persist outside proof blocks (hence the
very similar documents in Figure~\ref{dag} and Figure~\ref{dag2}
have a different semantics).
This is what naturally
happens if the system has a single, global, imperative state that is updated
by the execution of commands. In our scenario the commands belonging to opaque
proofs may even be executed remotely, hence a global side effect is
lost. To preserve the current semantics of documents, when a global command is
found in the middle of a proof its corresponding transaction is placed in the
DAG twice: once in the proof branch at its original position, and another time
on the master branch. This duplication is truly necessary: only the
transaction belonging to the master branch will retain its global effect; the
one in the proof branch, begin part of a pure computation, will have a local
effect only. In other words the effect of the ``\lstinline/Hint/\ldots''
transaction from state 3 to 4
is limited to states 4 and 5.

Branches identifying opaque proofs also define compartments from which errors
do not escape. If a proof contains an error one can replace such proof by
another one, possibly a correct one, without impacting the rest of the
document. As long as the statement of an opaque proof does not change,
altering the proof does not require re-checking what follows it in the document.

\section{Modelling asynchronously computed proofs}\label{futures}

The logical environment of a prover contains the statements of the theorems
which proof has already been checked. In some provers, like Coq, it also
contains the proof evidence, a proof term. A theorem enters the logical
environment only if its proof evidence has been checked. In our scenario the
proof evidences may be produced asynchronously, hence this condition has to be
relaxed, allowing a theorem to be part of the environment before its proof is
checked.  Wenzel introduced the very general concept of proof
promise~\cite{Wenzel_parallelproof} in the core logic of Isabelle for this
precise purpose.

\subsection{Proof promises in Coq}

Given that only opaque proofs are processed asynchronously we can
avoid introducing in the theory of Coq the generic notion of a
sub-term asynchronously produced.  The special status of opaque
proofs spares us to change the syntax of the terms and lets us
just act on the logical environment and the Well Founded (WF)
judgement.  The relevant rules for the WF judgements, in
the presentation style of~\cite{compact}, are the following ones.
We took the freedom to omit some details, like $T$ being a well
formed type, that play no role in the current discussion.

\begin{center}
\AxiomC{$E \vdash$ WF $\quad E \vdash b : T \quad d$ fresh in $E$}
\UnaryInfC{$E \cup (\mbox{\textbf{definition }} d : T := b) \vdash$ WF}
\DisplayProof
~
\AxiomC{$E \vdash$ WF $\quad E \vdash b : T \quad d$ fresh in $E$}
\UnaryInfC{$E \cup (\mbox{\textbf{opaque }} d : T ~|~ b) \vdash$ WF}
\DisplayProof
\end{center}

Note that the proof evidence $b$ for opaque proofs is checked but
not stored in the environment as the body of non-opaque definitions.
After an opaque proof enters the environment, it shall behave exactly
like an axiom.  

\begin{center}
\AxiomC{$E \vdash$ WF $\quad d$ fresh in $E$ }
\UnaryInfC{$E \cup (\mbox{\textbf{axiom }} d : T) \vdash$ WF}
\DisplayProof
\end{center}

We rephrase the WF judgement as the combination of
two new judgements: Asynchronous and Synchronous Well Founded (AWF and
SWF respectively).
The former is used while the user interacts with the system.
The latter complements the former when the complete
checking of the document is required.

\begin{center}
\AxiomC{$E \vdash$ AWF $\quad E \vdash b : T \quad d$ fresh in $E$}
\UnaryInfC{$E \cup (\mbox{\textbf{definition }} d : T := b) \vdash$ AWF}
\DisplayProof
~
\vspace{1.2em}
\AxiomC{$E \vdash$ AWF $\quad d$ fresh in $E$}
\UnaryInfC{$E \cup (\mbox{\textbf{opaque }} d : T ~|~ [f]_E) \vdash$ AWF}
\DisplayProof

\AxiomC{$E \vdash$ SWF}
\UnaryInfC{$E \cup (\mbox{\textbf{definition} } d \!:\! T := b ) \vdash$ SWF}
\DisplayProof
~
\vspace{1.2em}
\AxiomC{$E \vdash$ SWF $\quad b = $ run $f$ in $E \quad E \vdash b \!:\! T$}
\UnaryInfC{$E \cup (\mbox{\textbf{opaque} } d \!:\! T ~|~ [f]_E) \vdash$ SWF}
\DisplayProof

\AxiomC{$E \vdash $ AWF}
\AxiomC{$E \vdash $ SWF}
\BinaryInfC{$E \vdash $ WF}
\DisplayProof
\end{center}

The notation $[f]_E$ represents a pure computation in the environment
$E$ that eventually produces a proof evidence $b$.
The implementation of the two new judgements amounts to replace the
type of opaque proofs (\lstinline/term/) with a function type
(\lstinline/environment -> term/) and impose that such computation is
pure.
In practice the commitment of Coq to not use the proof terms of
theorems terminated with \tmverbatim{Qed} makes it possible to run
these computations when it is more profitable. In our implementation
this typically happens in the background.

As anticipated the Coq system is coded using the
imperative features provided by the OCaml language quite pervasively.
As a result we cannot simply rely on the regular
function type (\lstinline/environment -> term/)
to model pure computations, since the code
producing proof terms is not necessarily pure.  Luckily the
\tmverbatim{Undo} facility lets one backup the state of the system and
restore it, and we can use this feature to craft our own, heavyweight,
type of pure
computations.  A pure computation $c_0$ pairs a function $f$ with the
system state it should run in $s_0$ (that includes the logical
environment).  When $c_0$ is executed, the
current system state $s_c$ is saved, then $s_0$ in installed and $f$
is run.  Finally the resulting value $v$ and resulting state $s_1$
are paired in the resulting computation $c_1$, and the original state
$s_c$ is restored. We need to save $s_1$ only if the computation $c_1$
needs to be chained with additional impure code.  A
computation producing a proof is the result of chaining few impure
functions with a final call to the kernel's type checker that is a
\emph{purely functional} piece of code.  Hence the final system state is
always discarded, only the associated value (a proof term) is
retained.

The changes described in this section
enlarged the size of the trusted code base of Coq
by less than 300 lines (circa 2\% of its previous size).

\subsection{Parallelism in OCaml}

The OCaml runtime has no support
for shared memory and parallel thread execution, but provides
inter process communication facilities like sockets and data
marshaling.
Hence the most natural way to exploit parallel hardware is to split
the work among different worker processes.

While this approach imposes a clean message passing discipline, it may
clash with the way the state of the system is represented and stocked.
Coq's
global state is unstructured and fragmented: each software
module can hold some private
data and must register a pair of functions to take a snapshot and restore an
old backup to a central facility implementing the \tmverbatim{Undo}
command.
Getting a snapshot of the system state is hence possible, but marshalling it
and sending it to a worker process is still troublesome. In particular the
system state can contain a lot of unnecessary data, or worse data that
cannot be sent trough a channel (like a file descriptor) or even data one does
not want to send to a precise worker, like the description of the tasks he is
not supposed to perform.

Our solution to this problem is to extrude from the system state the unwanted
data, put a unique key in place of it and finally associate via a separate
table the data to the keys. When a
system state is sent to a worker process the keys it contains lose validity,
hence preventing the worker process to access the unwanted data.

The only remaining problem is that, while when a system state becomes
unreferenced it is collected by the OCaml runtime, the key-value table still
holds references to a part of that system state. Of course we want the
OCaml runtime to also collect such data. This can be achieved by making the
table ``key-weak'', in the sense that the references it holds to its keys do
not prevent the garbage collector from collecting them and that when this
happens the corresponding data has also to be collected. Even if the OCaml
runtime does not provide such notion of weak table natively, one can easily
code an equivalent finalization mechanism
known as \emph{ephemeron}
(key-value) pairs~\cite{Hayes:1997:ENF:263698.263733} 
by attaching to the keys a custom finalization function.

\begin{figure}[!h]
\begin{center}
\includegraphics[width=0.7\textwidth]{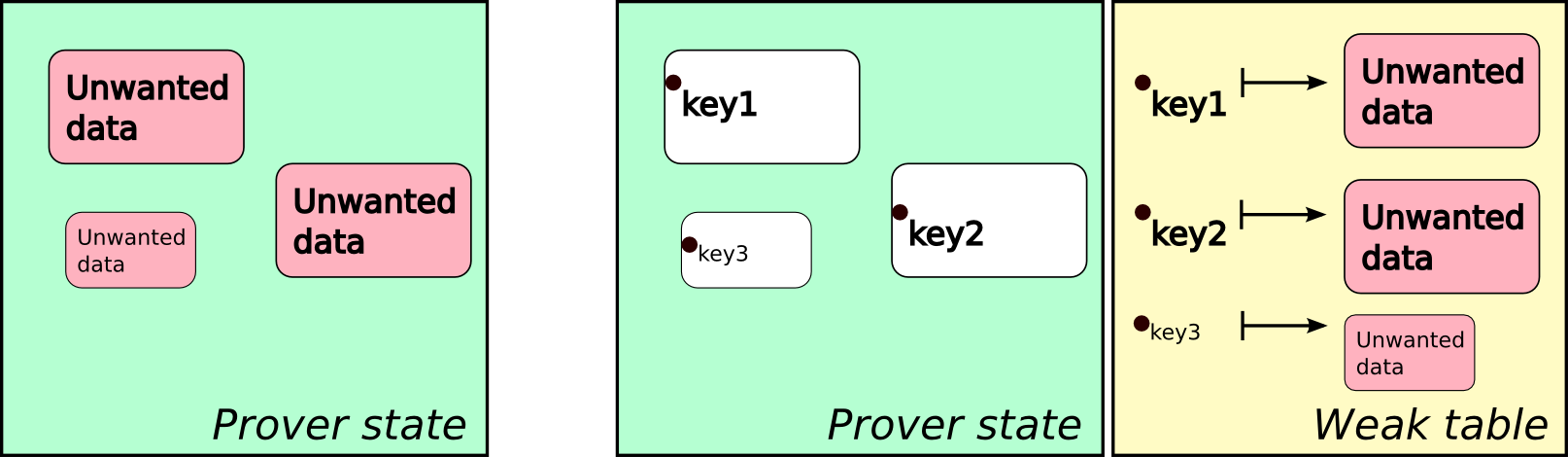}
\end{center}
\caption{Reorganization of the prover state}
\end{figure}

\subsection{The asynchronous task queue and the quick compilation
chain}\label{sepcomp}

The STM maintains a structured
representation of the document the prover is processing, identifies
independent tasks and delegates them to worker processes.
Here we focus on the interaction between Coq and the pool of
worker processes. 

The main kind of tasks identified by the STM is the one of opaque proofs we
discussed before, but two other tasks are also supported. The first one is
queries: commands having no effect but for producing a human readable output.
The other one is tactics terminating a goal applied to a set of goals via the
``\tmverbatim{par:}'' goal selector.{\footnote{The ``\tmverbatim{par:}'' goal
selector is a new feature of Coq 8.5 made possible by this work}} In all cases
the programming API for handling a pool of worker processes is the
\tmverbatim{AsyncTaskQueue},
depicted Figure~\ref{asyncqueue}. Such data structure, generic
enough to accommodate the three aforementioned kind of tasks, provides a priority
queue in which tasks of the same kind can be enqueued. Tasks will eventually
be picked up by a worker manager (a cooperative thread in charge of a
specific worker process), turned into requests and sent to the
corresponding worker.

\begin{figure}[h]
\begin{lstlisting}
module type Task = sig
  type task
  type request
  type response
  val request_of_task : [ `Fresh | `Old ] -> task -> request option
  val use_response : task -> response -> [ `Stay | `Reset ]
  val perform : request -> response
end
module MakeQueue(T : Task) : sig
  val init : max_workers:int -> unit
  val priority_rel : (T.task -> T.task -> int) -> unit
  val enqueue_task : T.task -> cancel_switch:bool ref -> unit
  val dump : unit -> request list
end
module MakeWorker(T : Task) : sig
  val main_loop : unit -> unit
end
\end{lstlisting}
\caption{\label{asyncqueue}Programming API for asynchronous task queues}
\end{figure}

By providing a description of the task, i.e. a module implementing the
\tmverbatim{Task} interface, one can obtain at once the corresponding queue
and the main function for the corresponding workers. While the
\tmverbatim{task} datatype can, in principle, be anything,
\tmverbatim{request} and \tmverbatim{response} must be marshalable since they
will be exchanged between the master and a worker processes.
\tmverbatim{request\_of\_task} translates a \tmverbatim{task} into a
\tmverbatim{request} for a given worker. If the worker is an \tmverbatim{`Old}
one, the
representation of the request can be lightweight, since the worker can save
the context in which the tasks take place and reuse it (for example
the proof context is the same for each goal to which the 
\tmverbatim{par:} goal selector applies). Also a \tmverbatim{task},
while waiting in the 
queue, can become obsolete, hence the option type. The worker
manager calls \tmverbatim{use\_response} when a response is received, and
decides whether the worker has to be reset or not. The \tmverbatim{perform}
function is the code that is run in the worker. Note that when a task is
enqueued a handle to cancel the execution if it becomes outdated
is given. The STM will flip that bit eventually, and the worker manager will
stop the corresponding worker.

This abstraction is not only valuable because it hides to the programmer all
the communication code (socket, marshalling, error handling) for the three
kinds of tasks. But also because
it enables a new form of separate {\emph{quick compilation}} (batch document
checking) that splits such job into two steps. A first and quick one that
essentially amounts at checking everything but the opaque proofs and
generates an (incomplete) compiled file, and a second one that completes
such file. An incomplete file,
extension \tmverbatim{.vio}, can already be used: as we pointed out before the
logic of Coq commits not to use opaque proofs, no matter if they belong to the
current file or to another one. 
Incomplete \tmverbatim{.vio} files can be completed into the usual
\tmverbatim{.vo} files later on, by resuming the 
list of requests also saved in the
\tmverbatim{.vio} file.  

The trick is to set the number of workers to zero when the queue is
initialized. This means that no task is processed at all when the document is
checked. One can then \tmverbatim{dump} the contents of the queue in terms of
a
list of requests (a marshalable piece of data) and stock it in
the  \tmverbatim{.vio} files.  Such lists represents all the tasks
still to be performed in order to check the opaque proofs of the document. 
The performances of the quick compilation chain are assessed in
Section~\ref{12core}.

\section{The user side}\label{pide}

A prover can execute commands out-of-order only if the user can
interact with it asynchronously. We need an interface that
systematically gives to the prover the document the user is working on
and declares where the user is looking at to help the system take
scheduling decisions.  This way the UI also frees the user
from the burden of explicitly sending a portion of its text buffer to
the prover.  Finally this interface needs to be able to interpret the
reports the prover generates in no predictable order, and display them
to the user.

An UI like this one makes the user experience in editing a formal
document way closer to the one he has when editing a regular document
using a mainstream word processor: he freely edits the text while the
system performs ``spell checking'' in the background, highlighting in
red misspelled words. In our case it will mark in red illegal
proof steps.  Contextual information, like the current goal being
proved, is displayed according to the position of the cursor.  It is
also natural to have the UI aggregate diverse kinds of feedbacks on
the same piece of text.  The emblematic example is the full names of
identifiers that are displayed as an hyperlink pointing to the place
where the identifier is bound.

\subsection{PIDE and its document model}

\begin{figure}[h]
  \begin{center}
  \includegraphics[width=0.9\textwidth]{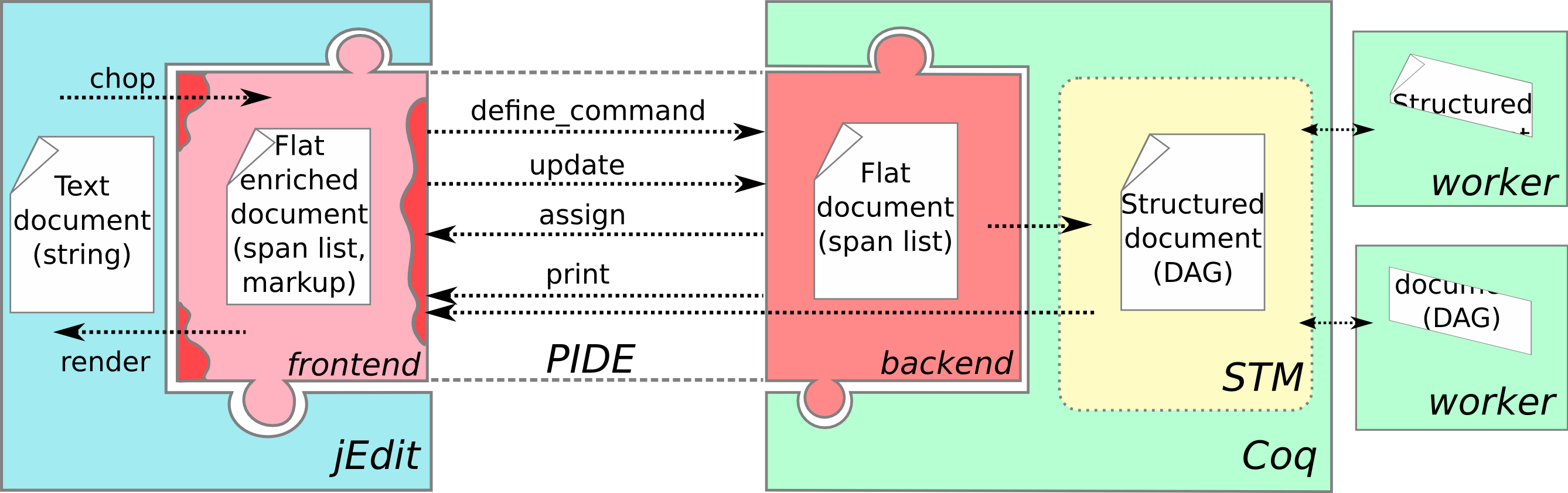}
  \end{center}
  \caption{PIDE sitting between the prover and the UI\label{fig:Architecture}}
\end{figure}

To integrate user interfaces with asynchronous provers in a uniform
way, Wenzel developed the PIDE middleware~{\cite{DBLP:journals/corr/Wenzel13}}.
This middleware consists of a number of API functions in a frontend
written in the Scala programming language,
that have an effect on a prover backend. The front-, and backend together
maintain a shared data structured, PIDE's notion of a \emph{document}.
Figure~\ref{fig:Architecture} depicts how the front and backend collaborate
on the document, and its various incarnations. The portions of the figure in
dark red represent parts of PIDE that were implemented to integrate Coq into
the PIDE middleware.

This document has a very simple, prover-independent, structure: it is a flat
list of spans.  A span is the smallest portion of text in which the
document is chopped by a prover specific piece of code.  In the case of Coq it
corresponds to a full command.
The frontend exposes an \emph{update} method on the
document, which allows an interface to notify the prover of changes. The
second functionality of the document on the frontend is the ability to query
the document for any markup attached to the command spans. This markup is
provided by the prover, and is the basis for the rich feedback we describe in
Section~\ref{sec:RichFeedback}.

The simple structure of the document keeps the frontend ``dumb'', which makes it
easy to integrate a new prover in PIDE: for Coq, this only required
implementing a parser that can recognize, chop, command spans (around
120 lines of Scala code) and a way to start the
prover and exchange data with it (10 more lines). 
The prover side of this communication is further
described in Section~\ref{sec:PIDETop}, which also describes the
representation of the document text in the PIDE backend for Coq.

The way an interface uses the PIDE document model is by updating it with new
text, and then reacting to new data coming from the prover, transforming it in
an appropriate way. For example, Coq can report the abstract syntax tree (AST)
of each command in the proof document, which the interface can use to provide
syntax highlighting which is not based on error-prone regular expressions.
Beyond the regular interface services, the information can also be used to
support novel interactions with a proof document. In the case of an AST, for
example, the tree can be used to support refactoring operations
robustly.

\subsection{Pidetop}\label{sec:PIDETop}

Pidetop is the toplevel loop that runs on the Coq side and translates
PIDE's protocol messages into instructions for the STM. Pidetop
maintains an internal representation of the shared document. When its
update function is invoked, the toplevel updates the internal document
representation, inserting and deleting command phrases where
appropriate.  The commands that are considered as new by pidetop are
not only those that have changed but also those that follow: because
the client is dumb, it does not know what the impact of a change is.
Instead it relies on the statical analysis the STM performs to
determine this.

Finally PIDE instructs the STM to start checking the updated document.
This instruction can include a \emph{perspective}: the set of
spans that frontend is currently showing to the user. This perspective
is
used by the STM to prioritize processing certain portions of the
document.

Processing the document
produces messages to update PIDE
of new information. During the computation, PIDE can interrupt the STM at any
moment, to update the document following the user's changes.

\subsection{Metadata collection}

The information Coq needs to report to the UI can be classified
according to its nature: persistent or volatile.  \emph{Persistent
information} is part of the system state, and can be immediately
accessed.  For this kind of information we follow the model of
asynchronous prints introduced by Wenzel in~\cite{Wenzel14}. When a
system state is ready it is reported to (any number of) asynchronous
printers, which process the state, reporting some information from it
as marked up messages for the frontend.  Some of these printers, like
those reporting the goal state at each span, are always executed,
while others can be printed on-demand. An example of the latter are
Coq's queries, which are scheduled for execution whenever the user
requests it.
\emph{Volatile information} can only be reported during the execution
of a command, since it is not retained in the system state.  An example
is the resolution of an identifier resulting in an hyper link: the
relation between the input text and the resulting term is not stored,
hence localized reports concerning sub terms can only be generated
while processing a command.  Volatile information can also be produced
optionally and on demand by processing commands a second time, for example in
response to the user request of more details about a term.
From the UI perspective both kind of reports are asynchronously
generated and handled homogeneously.

\subsection{Rich Feedback}\label{sec:RichFeedback}

\begin{figure}
  \includegraphics[width=\textwidth]{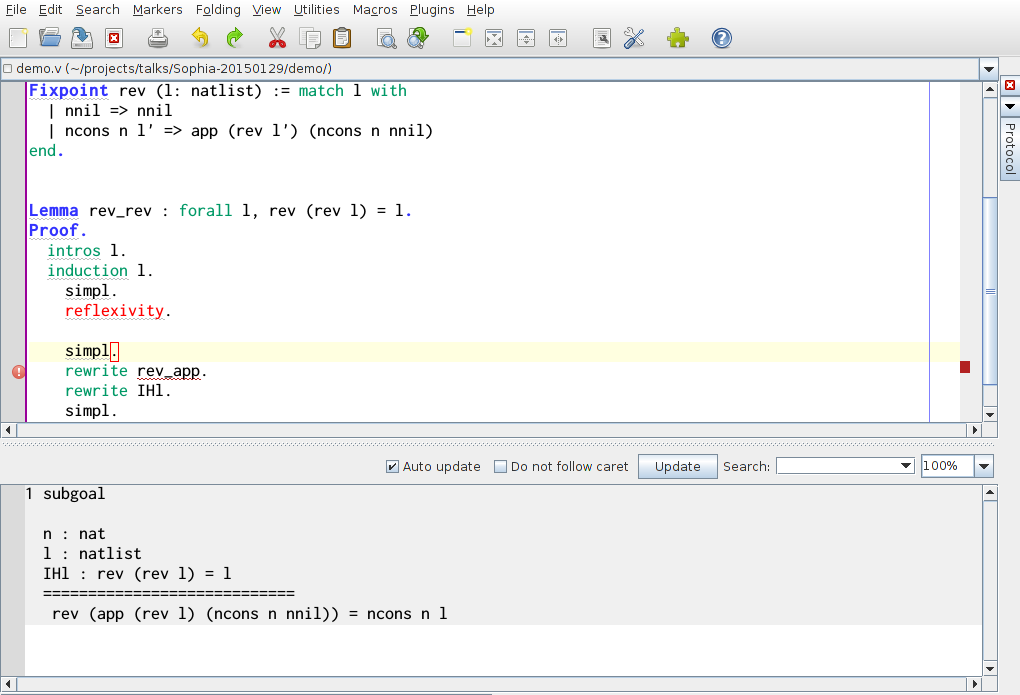}
  \caption{jEdit using PIDE \label{fig:jEdit-screenshot}}
\end{figure}

A user interface building on PIDE's frontend can use the marked up
metadata produced by the prover to provide a much richer editing
experience than was previously possible: the accessible information is no
longer restricted to
the output of
the last executed command.
For example jEdit, the reference frontend based on PIDE,
uses the spell checker idiom to report problems in the entire proof
document: instead of refusing to proceed beyond a line containing an
error the interface underlines all the offending parts in red, and the
user can inspect each of them.

This modeless interaction model also allows the interface to take
different directions than the write-and-execute one imposed by REPL
based UI.  For example it is now easy to decorrelate the user's point
of observation and the place where she is editing the document.  A
paradigmatic example concerns the linguistic construct of postfix
bookkeeping of variables and hypotheses (the \lstinline{as} intro
pattern in Coq).  The user is encouraged to decorate commands like the one
that starts an induction with the names of the new variables and
hypotheses that are available in each subgoal.  This decoration is
usually done in a sloppy way: when the user starts to work
on the third subgoal she goes back to the induction command in order
to write the intro pattern part concerning the third goal.  In the
REPL model she would lose the display of the goal she is actually working
on because the only goal displayed.
would be the one immediately
before the induction command.  By using a PIDE based interface to Coq
the user can now leave the observation point on the third goal and see
how it changes \emph{while} she is editing the intro pattern (note the
``Do not follow caret'' check box in
Figure~\ref{fig:jEdit-screenshot}).

These new ways of interaction are the first experiments with the PIDE
model for Coq, and we believe that even more drastically different
interaction idioms can be provided, for example allowing the user to
write more structured proofs, where the cases of a proof are gradually
refined, possibly in a different order than the one in which they
appear in the finished document. 

The PIDE/jEdit based interface for Coq 8.5 is distributed at the
following URL: \url{http://coqpide.bitbucket.org/}.

\section{Assessment of the quick compilation chain}\label{12core}

As described in Section~\ref{sepcomp} our design enables Coq to postpone the
processing of opaque proofs even when used as a batch compiler. To have a
picture of the improvement in the reactivity of the system we consider the time
that one needs to wait (we call that latency) in order to be able to use
(\tmverbatim{Require} in Coq's terminology) the last file of the proof
of the Odd Order Theorem after a change in the first file of such proof.
With Coq 8.4 there is nothing one can do but to wait for the full
compilation of the whole formalization. This takes a bit less than two and a
half hours on a fairly recent Xeon 2.3GHz machine using a single core (first
column). Using two cores, and compiling at most two files at a time, one can
cut that time to ninety minutes (second column). Unfortunately batch
compilation has to honour the dependency among files and follow a topological
order, hence by adding extra 10 cores one can cut only twelve minutes (third
column). Hence, in Coq 8.4 the best latency
is of the order of one hour.

\begin{figure}[h]
  \includegraphics[width=\textwidth]{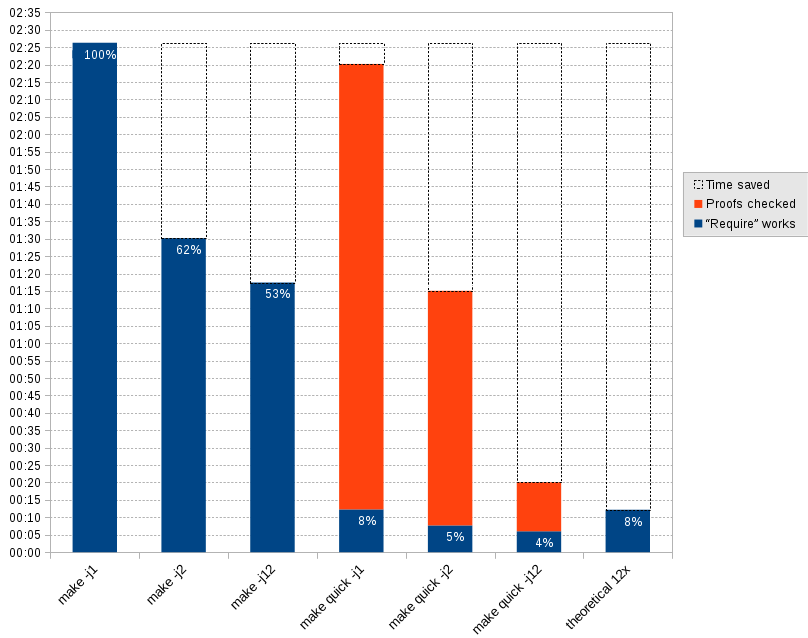}
  \caption{Benchmarks}
\end{figure}

Thanks to to the static analysis of the document described in
Section~\ref{static}, the work of checking
definitions and statements (in blue) can be separated by the checking of 
opaque proofs
(in red), and one can use a (partially) compiled Coq file even if its opaque
proofs are not checked yet.
In this scenario, the same hardware gives a latency of twelve minutes
using a single core (fourth column), eight minutes using two cores (fifth
column) and seven minutes using twelve cores (sixth column). After that time
one can use the entire formalization. 
When one then decides to complete the compilation he can
exploit the fact that each proof is independent to obtain a
good degree of parallelism. For example
checking all proofs using twelve cores requires thirteen extra minutes after
the
initial analysis, for a total of twenty minutes. This is 166\% of the
theoretical optimum one could get (seventh column). Still with
12 cores the latency is
only six minutes, on par with the theoretical optimum for 24 cores.

 The reader may notice that the quick compilation chain using 1 core (4th
 column) is slightly faster than the standard compilation chain. 
 This phenomenon concerns only the
 largest and most complicated files of the development.
 To process these files Coq requires a few gigabytes of
 memory and it stresses the OCaml garbage collector quite a bit (where it spends
 more than 20\% of the time). The separation of the two compilation phases
 passes trough marshalling
 to (a fast) disk and un-marshaling to an empty process space. This operation
 trades (non blocking, since the size of files fits the memory of the computer)
 disk I/O for a more compact and less fragmented memory layout that makes the
 OCaml runtime slightly faster.

\section{Concluding remarks and future directions}\label{conclusions}

This paper describes the redesign Coq underwent in order to provide a
better user experience, especially when used to edit large formal
developments.  The system is now able to better exploit parallel
hardware when used in batch mode, and is more reactive when used
interactively. In particular it can now talk with user
interfaces that use the PIDE middleware, among which we find the one
of Isabelle~\cite{Isabelle} that is based on jEdit and the
Coqoon~\cite{coqoon,coqoon2} one based on Eclipse.

There are many ways this work can be improved.
The most interesting paths seem to be the following ones.

First, one could make the prover generate, on demand, more metadata for the
user consumption.  A typical example is the type of sub expressions to
ease the reading of the document.  Another example is the
precise list of theorems or local assumptions used by automatic tactics like
\tmverbatim{auto}. This extra metadata could be at the base of assisted
refactoring functionalities the UI could provide.

Another interesting direction is to refine the static analysis of the
document to split proofs into smaller, independent, parts.  In a
complementary way one could make such structure easier to detect in
the proof languages supported by the system.  The extra structure
could be used in at least two ways.
First, one could give more accurate
feedback on broken proofs. Today the system stops at the first error it
encounters, but a more precise structure would enable the system to
backup by considering the error confined to the sub-proof in which it
occurs.
Second, one could increase the degree of parallelism we can exploit.

Finally one could take full profit of the PIDE middleware by adapting to
Coq interesting user interfaces based on it.  For example
clide~\cite{RingL14} builds on top of PIDE
and provides a web based, collaborative, user interface for the
Isabelle prover.  The cost of adapting it to work with Coq seems
now affordable.

\bibliographystyle{abbrv}

\bibliography{generic}

\end{document}